\documentclass[11pt, a4paper]{article}

\usepackage[margin=2.5cm]{geometry} 
\usepackage{setspace}
\usepackage[utf8]{inputenc}
\usepackage[T1]{fontenc}

\usepackage{amsmath, amssymb, amsthm}
\usepackage{mathtools}
\usepackage{titlesec}
\titleformat*{\section}{\large\bfseries}
\titleformat*{\subsection}{\normalsize\bfseries}
\titleformat*{\paragraph}{\normalsize\bfseries}

\usepackage{graphicx}
\usepackage{caption}
\usepackage{subcaption}
\usepackage[table]{xcolor}
\usepackage{booktabs}
\usepackage{threeparttable}
\usepackage{siunitx}
\usepackage{etoolbox} % このパッケージが必要です
\robustify\bfseries   % S列の中で\bfseriesを効かせるためのおまじない
\usepackage{multirow}
\usepackage{tabularx}

\usepackage{ragged2e}

\newcolumntype{Y}{>{\centering\arraybackslash}X} % ←均等幅列が必要な場合のみ

\usepackage[round,authoryear]{natbib}
\usepackage[colorlinks=true,allcolors=blue]{hyperref}
\usepackage{authblk}
\theoremstyle{plain}

\theoremstyle{definition}

\usepackage{titlesec}

\titleformat{\paragraph}[runin]
  {\normalfont\slshape}   % 太字にしない
  {\theparagraph}
  {1em}
  {}              % 見出し本体
  [. ]            % 見出し後の区切り（ピリオド＋空白）

\usepackage{enumitem}
\setlist[description]{
  leftmargin=0pt,     % 左にはみ出さない
  labelindent=0pt,    % インデントなし
  itemsep=0.1em       % 項目間の空き
}

\title{\large \textbf{The Impact of Dodd-Frank and the Huawei Shock \\ on DRC Tin Exports}}

\author[1]{Haruka Nagamori\thanks{Email: u62501m@m.chukyo-u.ac.jp}}
\author[2]{Kazuhiko Nishimura\thanks{Email: nishimura@lets.chukyo-u.ac.jp}}
\affil[1]{Graduate School of Humanities and Social Sciences, Chukyo University, 466-8666 Japan}
\affil[2]{Institute of Economics, Chukyo University, 466-8666 Japan}

\date{\small \today} % 日付も少し小さく

\begin{document}

\maketitle

\begin{abstract}
\noindent
This paper investigates the structural transformation of the Democratic Republic of the Congo (DRC) tin market induced by the U.S. Dodd-Frank Act. 
Focusing on the breakdown of the pricing mechanism, we estimate the price elasticity of export demand from 2010 to October 2022 using a structural identification strategy that overcomes the lack of reliable unit value data. 
Our analysis reveals that the regulation effectively destroyed the price mechanism, with demand elasticity dropping to zero. 
This indicates the formation of a ``captive market'' driven by certification requirements rather than price competitiveness. 
Also, we find strong hysteresis; deregulation alone failed to restore market flexibility. 
The structural rigidity was finally broken not by policy suspension, but by the 2019 ``Huawei shock,'' an external demand surge that forced supply chain diversification. 

\vspace{0.5cm}
\noindent \text{Keywords:} Conflict minerals; Dodd-Frank Act; Trade elasticity; Market structure; Hysteresis; DRC
\end{abstract}

\clearpage
%\end{document}
% ==============================================================================
% MAIN CONTENT
% ==============================================================================

\section{Introduction}

The link between natural resource extraction and armed conflict---the so-called resource curse---has long been a central concern in development economics and international policy. In an effort to sever this link, Section 1502 of the Dodd-Frank Wall Street Reform and Consumer Protection Act (Dodd-Frank Act) was enacted in the United States in 2010. This landmark regulation required U.S.-listed companies to disclose their use of conflict minerals (tin, tantalum, tungsten, and gold; 3TG) originating from the Democratic Republic of the Congo (DRC) and adjoining countries.

While the policy aimed to promote responsible sourcing, a growing body of empirical literature suggests it generated severe unintended consequences. Seminal studies have documented a de facto boycott of DRC minerals that exacerbated infant mortality \citep{Parker2016} and potentially fueled local violence \citep{Stoop2018}. \citet{Koch2018} critically reviewed these competing narratives, highlighting that while the ``de facto embargo'' narrative has been dominant, trade volumes eventually showed signs of recovery. Indeed, \citet{Schutte2019} analyzed trade flows of tantalum and tin, documenting how due diligence normalized logistics and reduced smuggling, suggesting a stabilization in \textit{volumes}.

However, focusing solely on volume masks the fundamental structural transformation of the market. As \citet{Freedman2011} noted during the regulation's inception, there was high hope that private sector certification schemes (such as iTSCI) would resolve the crisis where state governance failed. Yet, the economic cost of this market segmentation remains understudied.
Recently, \citet{Hanai2021} investigated the mechanism changes induced by the regulation, arguing that while actors' behaviors changed, the underlying political economy mechanisms of exploitation persisted.
These existing studies predominantly focus on trade volumes, transparency, or political dynamics, however.
What is missing is a quantitative analysis of the \textit{economic pricing mechanism}---specifically, whether DRC exporters act as competitive price takers or face a distorted, rigid demand structure.
Did the regulation merely reduce exports, or did it transform the DRC from a competitive supplier into a captive source locked into specific supply chains?

This paper addresses these questions by estimating the price elasticity of export demand for DRC tin from 2010 to October 2022.
To ensure the integrity of our analysis and account for transit trade and potential misreporting, we integrate mirror trade data from both the DRC and the neighboring Republic of the Congo (Brazzaville).
Focusing on tin is particularly pertinent as it is the most economically significant conflict mineral for the region.
A major empirical challenge in this context is the lack of reliable export price data (unit values) during the regulatory turmoil.
To overcome this, we adopt the structural identification strategy proposed by \citet{NakanoNishimura2025}, originally developed for service trade.
By exploiting bilateral exchange rate variations and deriving structural parameters from reduced-form estimates, we recover the demand elasticity even when price data are unobservable.

Our approach departs from standard Difference-in-Differences (DID) frameworks often used in policy evaluation.
As we document in Section 2, the Dodd-Frank regulation triggered a drastic, endogenous switching of trading partners---from a diverse portfolio led by China to a near-total concentration on certified smelting hubs such as Malaysia and Thailand---which invalidates the stable control group assumption required for DID.
Instead, we employ a structural break analysis, estimating the elasticities separately for four distinct regimes: (1) Pre-regulation, (2) The Dodd-Frank implementation, (3) The suspension period, and (4) The post-Huawei ban recovery era.

We present two key findings. First, the regulation destroyed the price mechanism.
Prior to the rule, DRC tin exports were highly price-elastic ($\eta \approx -27.5$), consistent with a competitive small open economy. 
However, following the implementation of the Dodd-Frank rule, the demand elasticity effectively dropped to zero (statistically insignificant), despite a surge in exports through certified channels.
This implies the formation of a captive market where trade was determined by regulatory certification networks rather than price competitiveness.
This finding aligns with qualitative evidence by \citet{Hanai2021}, who noted that non-certified minerals were sold to Chinese buyers at deep discounts (30--60\%), indicating a bifurcation of the market.

Second, we identify a strong hysteresis effect and a unique mechanism of recovery.
The suspension of the rule in 2017 did not immediately restore market elasticity; the captive supply chain persisted.
Strikingly, it was not the deregulation itself, but a massive external shock---the so-called Huawei shock---that finally shattered this rigidity.
On May 15, 2019, the U.S. Department of Commerce placed Huawei Technologies on the Entity List.
This geopolitical upheaval triggered a panic-buying spree within the global electronics supply chain.
Reflecting the long-standing strategic Sino-European race for African minerals \citep{Ebner2015}, Chinese manufacturers aggressively scrambled to secure inventories of tin---a critical material for semiconductor soldering---bypassing the established, rigid certification channels.
This %overwhelming 
surge in demand forced a diversification of export destinations, effectively breaking the captive structure and restoring the market's sensitivity to price signals ($\eta \approx -10.4$).
%Importantly, our robustness checks confirm that this structural recovery was specifically triggered by the Huawei shock, rather than merely reflecting a substitution effect from the concurrent decline in Myanmar's tin production.

This study contributes to the literature by moving beyond the analysis on volumes to providing the first empirical evidence of structural rigidity induced by conflict mineral regulations.
Our findings offer a cautionary tale for policymakers: regulations can create monopolistic bottlenecks that are resilient to simple deregulation, requiring significant external shocks to reset the competitive landscape.

\section{Institutional Background and Data}

\begin{figure}[t!]
  \centering
  \includegraphics[width=1.0\textwidth]{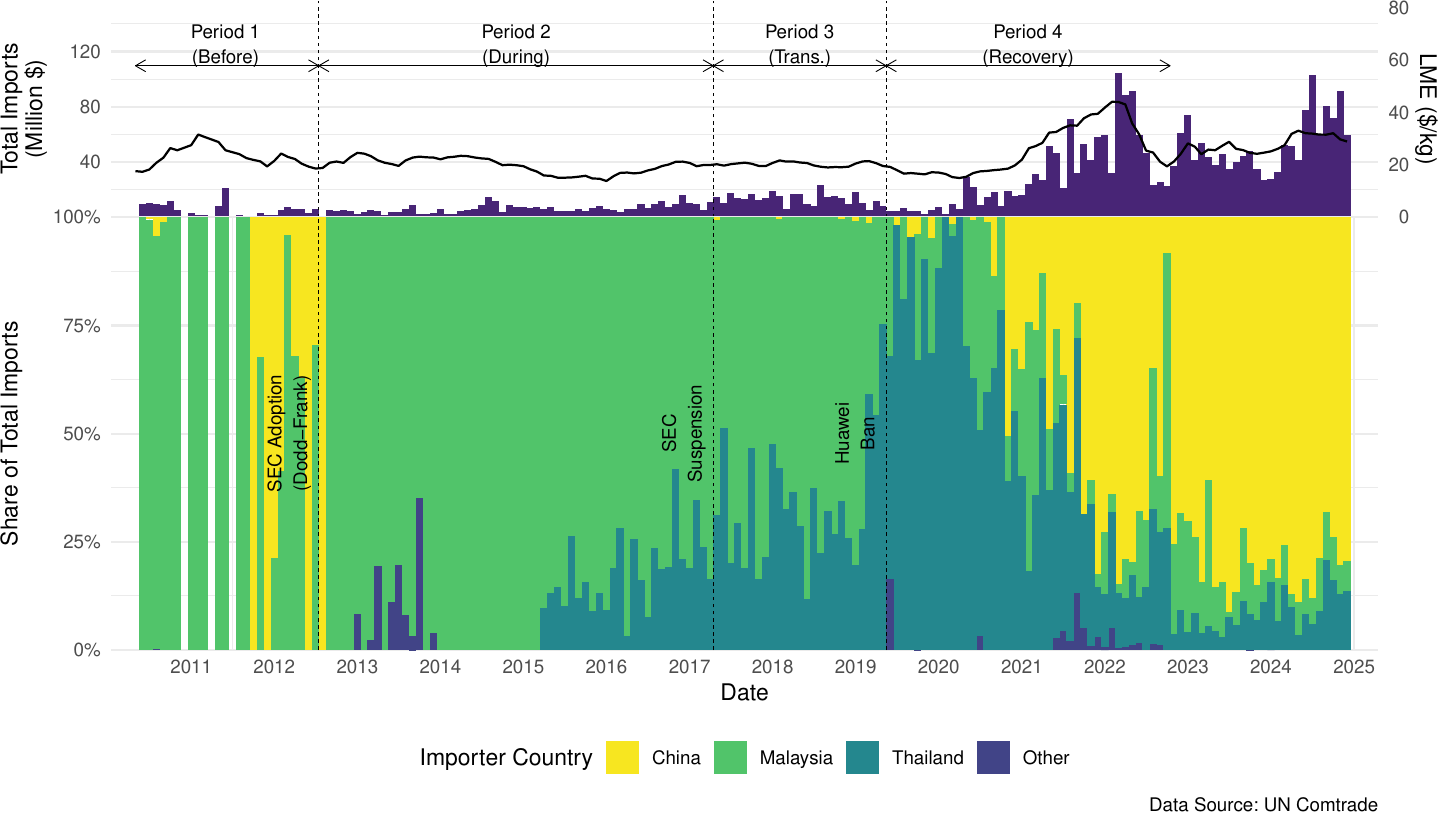}
  \caption{Export Destinations of DRC Tin and Global Tin Prices. The stacked bars represent the monthly total export value (million USD) and the share of importing countries (China in yellow, Malaysia in green, Thailand in teal, and others in dark blue). The solid black line indicates the LME tin price (USD/kg). The vertical dashed lines correspond to the adoption of the SEC conflict mineral rule (July 2012), the suspension of its enforcement (April 2017), and the U.S. ban on Huawei (May 2019). To account for transit trade and potential misreporting, data includes combined exports from both the DRC and the Republic of the Congo (Brazzaville). Source: Authors' calculation based on UN Comtrade and World Bank data.}
  \label{fig:export_share}
\end{figure}

\subsection{Timeline of Events}
To understand the impact of international regulations on the DRC tin market, we focus on three pivotal policy events that have fundamentally altered the trade landscape over the past decade.
Figure \ref{fig:export_share} illustrates the evolution of the share of importing countries for DRC tin, alongside global tin prices, from 2010 to 2025. The vertical dashed lines indicate the timing of these key events.
The timeline for our structural estimation is divided into four distinct periods:
\begin{description}
\item{\itshape Period 1: Before Regulation ($\le$ July 2012)}~~
Prior to the full implementation of the conflict mineral regulation, the export destinations of DRC tin were relatively diversified, with China (indicated by the yellow bars in Figure \ref{fig:export_share}) being a significant importer.
This period represents the baseline market structure characterized by competitive trading relationships.
\item{\itshape Period 2: The Dodd-Frank Era (August 2012 -- April 2017)}~~
In July 2010, the U.S. Congress passed the Dodd-Frank Wall Street Reform and Consumer Protection Act.
Section 1502 of the Act required U.S.-listed companies to disclose whether their products contained conflict minerals (tin, tantalum, tungsten, and gold) originating from the DRC.
The Securities and Exchange Commission (SEC) adopted the final rule in August 2012.
As Figure \ref{fig:export_share} clearly shows, the market structure shifted radically immediately following this adoption.
Exports became almost exclusively concentrated on certified smelting hubs in Malaysia (green bars) and Thailand (teal bars), while direct exports to China virtually vanished.
This drastic shift suggests the formation of a ``captive market,'' where DRC exporters were locked into specific supply chain routes certified by the conflict-free sourcing initiative (CFSI).
\item{\itshape Period 3: The Transition / Hysteresis (May 2017 -- May 2019)}~~
In April 2017, the SEC suspended the enforcement of the conflict mineral disclosure rule, effectively relaxing the regulatory pressure.
However, Figure \ref{fig:export_share} reveals a striking persistence in the trade pattern: despite the regulatory relief, the dominance of the Malaysian and Thai routes continued, and the market structure did not immediately revert to the pre-regulation state.
This observation points to a strong \textit{hysteresis} effect in the supply chain;
once established, the certified trade channels and commercial relationships were rigid and difficult to dismantle.
\item{\itshape Period 4: The Huawei Shock / Recovery (June 2019 -- October 2022)}~~
The turning point arrived in May 2019, when the U.S. Department of Commerce placed Huawei and its affiliates on the Entity List (the Huawei Ban).
This exogenous shock disrupted global semiconductor supply chains and forced Chinese manufacturers to aggressively secure tin resources from alternative sources.
Coinciding with this event, we observe a significant diversification in DRC export destinations.
The share of Malaysia and Thailand declined, and China re-emerged as a major importer.
This period marks the restoration of a more competitive market structure, driven not by internal regulatory changes but by an external demand shock.
\end{description}

Our structural estimation in this paper restricts Period 4 to October 2022. As shown in Figure \ref{fig:export_share}, late 2022 witnessed a sharp, simultaneous collapse in both trade volumes and LME prices. This marks a transition into a severe macroeconomic contraction phase driven by post-pandemic inventory adjustments and China's chaotic exit from the zero-COVID policy. We exclude this subsequent period to isolate the regulatory-driven market recovery from these overarching global macroeconomic shocks.

%\vspace{0.5cm}
This drastic and endogenous switching of trading partners---from a competitive mix to a concentrated buyer group, and then back to diversification---invalidates the parallel trend assumption required for standard Difference-in-Differences (DID) estimation.
Consequently, instead of a DID approach, we adopt a structural break analysis, estimating the trade elasticities separately for these four distinct regimes to capture the changing nature of market competition.

\subsection{Data Construction}

To empirically analyze the structural shifts in the DRC tin market, we construct a monthly panel dataset covering the period from May 2010 to October 2022. Our dataset integrates trade flows, bilateral exchange rates, and global commodity prices.

\begin{description}
\item{\itshape Trade Value Data}~~
We source monthly export data for tin ores and concentrates (HS Code 260900) from the UN Comtrade database.\footnote{We focus on tin rather than gold because the regulatory impact on market structure is more observable in the tin sector.
\citet{Hanai2021} points out that while the 3T (tin, tantalum, tungsten) sector saw a significant withdrawal of armed groups and progress in certification, the gold sector remains largely dominated by smuggling networks due to its high value-to-weight ratio and ease of illicit transport.} 
Given the well-documented irregularities in direct reporting by the DRC customs authorities during periods of conflict and regulatory turmoil, we rely on \textit{mirror statistics} (i.e., import data reported by the DRC's trading partners).\footnote{The reliability of trade data has significantly improved over time.
\citet{Schutte2019} compared regional export records with global import data and found that while large discrepancies existed prior to 2011 due to smuggling and misreporting, these discrepancies largely disappeared after 2012 (averaging 99\% consistency), coinciding with the expansion of due diligence schemes.}
To address potential misreporting and the laundering of minerals through neighboring transit countries, we aggregate the mirror trade data of the DRC with that of the Republic of the Congo (Brazzaville). Because the Republic of the Congo lacks commercial tin mining infrastructure, exports reported as originating from there are overwhelmingly laundered DRC tin \citep{Schutte2019}. Integrating these flows allows us to capture the true volume of DRC tin exports more accurately and ensures the robustness of our estimations against shifting smuggling routes.
\item{\itshape Price Data}~~
Data availability imposes two significant constraints on our analysis.
First, consistent monthly mirror data are only available from May 2010 onwards, preventing us from extending the pre-regulation baseline further back in time.
Second, and most critically, data on export \text{quantities} (and thus unit values) are largely missing prior to January 2017. This lack of price data for the majority of the sample period precludes the estimation of a standard structural demand system, thereby necessitating the reduced-form approach described in Section 3.
%Consequently, our dependent variable, $\ln X_{jt}$, is defined as the natural logarithm of the monthly export value (in current USD).
\item{\itshape Exchange Rates}~~
The bilateral exchange rate $E_{jt}$ is defined as the units of importer $j$'s currency per unit of DRC Franc (CDF).
An increase in $E_{jt}$ represents an appreciation of the importer's currency (or a depreciation of the CDF), which theoretically lowers the relative price of DRC exports.
To ensure comprehensive coverage of all trading partners---including emerging economies like Malaysia and Rwanda, which became pivotal during the study period---we sourced historical exchange rate data from \textit{FXTOP}, as standard sources such as FRED (Federal Reserve Economic Data) lacked coverage for several minor currencies in our sample.
We calculate the cross-rates using the monthly average exchange rates against the US Dollar.
\item{\itshape Global Market Price}~~
To control for global demand shocks and price cycles, we include the monthly average cash-settled price of Tin from the London Metal Exchange (LME), denoted as $\ln P_{t}^\text{LME}$.
This variable captures the exogenous baseline price movements common to all trading partners.
\item{\itshape Instrumental Variable Construction}~~
As discussed in Section 3, addressing the potential endogeneity of exchange rates requires a valid instrumental variable.
To implement this strategy, we construct an instrument based on the exchange rates of third countries.
Our procedure for selecting the optimal instrument for each partner currency $j$ is as follows:
\begin{enumerate}
    \item We computed pairwise correlations of monthly exchange rate fluctuations between currency $j$ and all other currencies in the global basket over the sample period.
    \item From the set of highly correlated currencies, we selected a third-country currency that belongs to a \textit{geographically distinct region} from country $j$.
\end{enumerate}
This geographic separation ensures that the instrument is correlated with the bilateral exchange rate (relevance) due to global financial trends but remains uncorrelated with idiosyncratic demand shocks specific to the trading partner's region (exclusion restriction).
For example, if the partner is an Asian country, we prioritize a highly correlated European or South American currency as the instrument.
\end{description}

\section{Identification Strategy}

\subsection{Structural Framework}
To estimate the price elasticity of demand for DRC tin exports, we consider a standard partial equilibrium model of trade.
Let $X_{jt}$ denote the value of tin exports from the DRC to country $j$ in month $t$.
The export value is defined as the product of the export quantity $Q_{jt}$ and the export price in the producer's currency $P_{t}^\text{exp}$.
Note that we assume the Law of One Price, implying that the export price $P_{t}^\text{exp}$ is common across all destination countries $j$ at time $t$.
The demand for DRC tin in country $j$, $Q_{jt}^\text{d}$, depends on the import price denominated in the local currency of country $j$, denoted as $P_{jt}^\text{imp}$.
The relationship between the producer's price and the importer's price is given by the bilateral exchange rate $E_{jt}$ (expressed as units of country $j$'s currency per unit of DRC currency) and iceberg transport costs $\tau_{j} \ge 1$:
\begin{equation}
    P_{jt}^\text{imp} = E_{jt} \cdot P_{t}^\text{exp} \cdot \tau_{j}
    \label{eq:price_def}
\end{equation}
where $\tau_{j}$ captures transport, insurance, and tariff costs (the wedge between CIF and FOB prices), which are assumed to be specific to each destination but constant over the estimation period.
Importantly, $\tau_j$ also captures destination-specific regulatory premiums or smuggling discounts.
Even if the market bifurcated into certified and non-certified channels during the regulatory era, these structural pricing differences are absorbed by $\tau_j$.
Furthermore, under our assumption of a highly elastic supply ($\omega \approx 0$, discussed below), the producer price $P_t^\text{exp}$ is pinned down by global market conditions rather than destination-specific demand shocks.
Consequently, potential violations of a strict Law of One Price across different supply chain routes do not bias our estimation, as time-invariant destination-specific wedges are fully absorbed by the importer fixed effects.

Assuming a constant elasticity of demand function, the demand equation is:
\begin{equation}
    \ln Q_{jt}^\text{d} = \alpha_j^0 + \eta \ln P_{jt}^\text{imp} + u_{jt}
    \label{eq:demand_qty}
\end{equation}
where $\eta$ is the price elasticity of demand (expected to be negative), $\alpha_j^0$ represents the base country-specific demand parameter, and $u_{jt}$ is the demand shock.
Since we rely on mirror statistics, our dependent variable is the import value reported by trading partners ($X_{jt} = P_{jt}^\text{imp} \cdot Q_{jt}$), which serves as a proxy for export value.
Thus, we rewrite Equation (\ref{eq:demand_qty}) in terms of this value.
Using $\ln X_{jt} = \ln P_{jt}^\text{imp} + \ln Q_{jt}$ and substituting Equation (\ref{eq:price_def}) into (\ref{eq:demand_qty}), we obtain:
\begin{equation}
    \ln X_{jt} = \alpha_j^0 + (1 + \eta) (\ln E_{jt} + \ln P_{t}^\text{exp} + \ln \tau_{j}) + u_{jt}
\end{equation}
The time-invariant transport cost term $(1+\eta)\ln \tau_{j}$ is naturally absorbed into the country fixed effects $\alpha_j$.
Thus, the structural demand equation in terms of value simplifies to:
\begin{equation}
    \ln X_{jt} = \alpha_j + (1 + \eta) (\ln E_{jt} + \ln P_{t}^\text{exp}) + u_{jt}
    \label{eq:demand_value}
\end{equation}
where $\alpha_j \equiv \alpha_j^0 + (1+\eta)\ln \tau_j$.

On the supply side, we first assume a standard constant-elasticity supply function depending on the producer price:
\begin{equation}
    \ln Q_{t}^s = \delta_0 + \varepsilon \ln P_{t}^\text{exp} + v_{t}
\end{equation}
where $\varepsilon$ is the price elasticity of supply (expected to be positive).
However, consistent with our estimation strategy, we specify the inverse supply function in terms of export value.
Using the identity $\ln X_{jt} = \ln P_{t}^\text{exp} + \ln Q_{jt}$, we rearrange the supply function as:
\begin{equation}
    \ln P_{t}^\text{exp} = \delta + \omega \ln X_{jt} + \tilde{v}_{t}
    \label{eq:supply_inverse}
\end{equation}
Here, the parameter $\omega$ is structurally defined as $\omega \equiv 1/(1+\varepsilon)$.
A value of $\omega \approx 0$ implies that $\varepsilon \to \infty$, consistent with the assumption of a perfectly elastic supply curve (i.e., the DRC is a price taker in the global tin market).

\subsection{Reduced Form Estimation}

A direct estimation of the structural demand equation (\ref{eq:demand_value}) faces a significant hurdle: the export price $P_{t}^\text{exp}$ is not consistently observed across all periods due to data limitations during the conflict-free regulation era.
Furthermore, $P_{t}^\text{exp}$ is endogenous, being simultaneously determined by the intersection of supply and demand curves.
This situation mirrors the challenge often encountered in service trade analysis, where price and quantity are inherently inseparable and unit values are unobservable.

To overcome this identification challenge, we adopt the structural approach proposed by \citet{NakanoNishimura2025}.
Originally developed to measure trade elasticities for services by exploiting exchange rate variations, this method allows for recovering structural parameters from reduced-form estimates.
Adopting this framework, we derive a reduced-form equation that relies solely on observable variables.
Substituting the inverse supply function (\ref{eq:supply_inverse}) into the structural demand equation (\ref{eq:demand_value}) allows us to eliminate the unobserved price term $P_{t}^\text{exp}$.
Substituting $\ln P_{t}^\text{exp} = \delta + \omega \ln X_{jt} + v_{t}$ into Equation (\ref{eq:demand_value}):
\begin{equation}
    \ln X_{jt} = \alpha_j + (1 + \eta) \left( \ln E_{jt} + (\delta + \omega \ln X_{jt} + v_{t}) \right) + u_{jt}
\end{equation}
Rearranging terms to solve for the dependent variable $\ln X_{jt}$:
\begin{equation}
    \ln X_{jt} \left(1 - \omega(1 + \eta) \right) = \text{Const.} + (1 + \eta) \ln E_{jt} + \text{Error}_{jt}
\end{equation}
Dividing both sides by $1 - \omega(1 + \eta)$, we obtain the following reduced-form specification:
\begin{equation}
    \ln X_{jt} = \mu_j + \beta \ln E_{jt} + \epsilon_{jt}
    \label{eq:reduced_form_final}
\end{equation}
where $\mu_j$ denotes the country fixed effects, and $\epsilon_{jt}$ is the composite error term.
Crucially, the estimated coefficient on the exchange rate, $\beta$, is a composite parameter defined as:
\begin{equation}
    \beta = \frac{1 + \eta}{1 - \omega(1 + \eta)}
    \label{eq:beta_definition}
\end{equation}
Our primary interest is to recover the true demand elasticity $\eta$ from the estimated $\beta$.
Equation (\ref{eq:beta_definition}) clarifies how the supply parameter $\omega$ affects this identification.
If the DRC is a small open economy with a perfectly elastic supply curve (i.e., price taker), then $\omega = 0$.
In this limiting case, the denominator becomes unity, and the reduced-form coefficient directly reveals the structural demand parameter:
\begin{equation}
    \lim_{\omega \to 0} \beta = 1 + \eta
\end{equation}

For a perfectly homogeneous commodity exported by a small open economy, the theoretical price elasticity of demand approaches $-\infty$.
In this light, an extremely large estimated elasticity (such as the $\eta \approx -27.5$ observed in Period 1) is not an anomaly;
rather, it is strong empirical evidence that the DRC was fully integrated into a highly competitive global market prior to the regulation.

If the supply curve is not perfectly horizontal ($\omega > 0$), we must consider the potential bias.
Since the demand elasticity $\eta$ is expected to be negative (and typically $\eta < -1$ for elastic goods), the term $(1+\eta)$ is negative.
Consequently, with $\omega > 0$, the denominator in Equation (\ref{eq:beta_definition}) becomes greater than one:
\[
\omega>0,\quad 1+\eta<0
\quad\Rightarrow\quad
1-\omega(1+\eta) > 1.
\]
This implies that the absolute value of our estimated coefficient $|\beta|$ will be smaller than the absolute value of the true structural parameter $|1 + \eta|$.
\begin{equation}
\left| \beta \right| = \left| \frac{1+\eta}{1 - \omega (1+\eta)} \right| < \left| 1+\eta \right|
\end{equation}
Therefore, assuming $\omega = 0$ provides a \textit{conservative lower bound} (in absolute terms) for the demand elasticity.
In other words, the true market response is likely even more elastic than our estimates suggest.
While supply conditions---such as conflict intensity or certification costs---may plausibly vary across regulatory regimes, our overall sample estimation yields an inverse supply elasticity that is statistically indistinguishable from zero ($\omega \approx 0.049$, $p=0.118$; see Appendix A).
Thus, we confidently proceed with the assumption $\omega \approx 0$ and interpret $\beta$ as $1+ \eta$.

Finally, to estimate Equation (\ref{eq:reduced_form_final}), we control for global market conditions by including the log of the LME tin price ($\ln P_{t}^\text{LME}$).
Since the bilateral exchange rate $\ln E_{jt}$ may still suffer from endogeneity due to omitted common shocks, we employ a Fixed Effects Instrumental Variables (FE-IV) estimator, using the exchange rates of third-country trading partners as instruments, following the strategy of \citet{Berman2012}.
A potential concern during periods of geopolitical stress, such as the Huawei shock, is that third-country exchange rates might correlate with global financial cycles or semiconductor demand surges, threatening the exclusion restriction.
However, global macro-financial dynamics and commodity super-cycles are instantaneously priced into the global benchmark.
By explicitly controlling for the LME tin price ($\ln P_{t}^\text{LME}$), we effectively absorb these common global demand shocks.
Consequently, our IV strategy isolates the residual, purely bilateral relative price variations, ensuring the validity of the instrument.

\section{Empirical Results}

Table \ref{tab:one} presents the estimation results of the reduced-form equation defined in Section 3. We estimate the model separately for the four periods identified in Section 2 to capture structural breaks in the trade regime.
All columns employ the Fixed Effects Instrumental Variables (FE-IV) estimator with Heteroskedasticity and Autocorrelation Consistent (HAC) standard errors.

%--------------------------------
\begin{table}[t!]
\centering
\begin{threeparttable}
\caption{Estimation of Export Demand Elasticity over Distinct Periods}
\label{tab:one}

% S列の設定: 整数部3桁、小数部3桁を確保。カッコとアスタリスクを許容
\begin{tabular*}{\textwidth}{@{\extracolsep{\fill}} 
  l 
  *{4}{S[
    table-format=-2.3, 
    table-space-text-post={***}, 
    input-symbols={()}, 
    table-align-text-post=false
  ]}
}
\toprule
& \multicolumn{4}{c}{Dependent Variable: $\ln{X}_{jt}$ (log of export value)} \\
\cmidrule(lr){2-5}
& {Period 1} & {Period 2} & {Period 3} & {Period 4} \\
& {Before Reg.} & {Dodd-Frank} & {Transition} & {Huawei Shock} \\
\midrule

\multicolumn{5}{l}{\textsl{Reduced Form Estimates}} \\
$\ln E_{j0,t}$ ($\beta$)
& -26.493{***} & 1.570   & 0.855   & -9.407{***} \\
& (4.164)      & (1.407) & (1.013) & (2.535)     \\
\addlinespace
$\ln P_{t}^{\text{LME}}$
& -4.916{***}  & 1.418   & -0.743  & 0.605   \\
& (1.078)      & (0.921) & (1.599) & (0.483) \\

\midrule
\multicolumn{5}{l}{\textsl{Implied Demand Elasticity (assuming $\omega \approx 0$)}} \\
$\eta$ ($= \beta - 1$)
& -27.493{***} & 0.570   & -0.145  & -10.407{***} \\
& (4.164)      & (1.407) & (1.013) & (2.535)      \\

\midrule
\multicolumn{5}{l}{\textsl{Diagnostics}} \\
Observations         & {39}    & {91}    & {55}    & {135}   \\
Number of Importers  & {3}     & {4}     & {3}     & {8}     \\
First-stage F-stat. & {153} & {17} & {265} & {28} \\
Hansen J p-val.    & 0.600   & 0.118   & 0.495   & 0.969   \\
\bottomrule
\end{tabular*}

\begin{tablenotes}[para,flushleft]
\footnotesize
\textit{Notes}: This table reports the instrumental variable (IV) fixed-effects panel estimations. The dependent variable is the log of monthly export value ($\ln V_{jt}$). Standard errors are robust to heteroskedasticity and autocorrelation (HAC) and reported in parentheses. Significance levels: * $p < 0.1$, ** $p < 0.05$, *** $p < 0.01$. Instrumental variables are the exchange rates of third-country trading partners' currencies. The first-stage F-statistic is the Kleibergen-Paap rk Wald F statistic.
\end{tablenotes}

\end{threeparttable}
\end{table}
%--------------------------------

\subsection{Model Validity and Diagnostics}
Before discussing the elasticities, we briefly assess the validity and robustness of our estimations.
The lower section of Table \ref{tab:one} reports the first-stage diagnostics. Despite the relatively small sample sizes in some periods, the Kleibergen-Paap rk Wald F-statistics indicate strong identification.
For Periods 1, 3, and 4, the F-statistics (153, 265, and 28, respectively) far exceed standard critical values.
Even in Period 2, the F-statistic of 17 surpasses the Stock-Yogo critical value for 10\% maximal IV size (19.93 for one endogenous variable and two instruments is approximately 19.9, but robustly passes the 15\% threshold of 11.59), mitigating concerns about weak instruments.
The coefficient on the LME tin price is negative and significant in Period 1, which may reflect specific pre-regulation volatility or unobserved supply disruptions. In Period 4, it turns positive (0.605), consistent with standard theory where global demand shocks drive exports, although our primary focus remains on the exchange rate elasticity.

Regarding finite-sample inference, we acknowledge that the estimations for specific regimes (particularly Periods 2 and 3) rely on relatively small samples in terms of both time periods and the number of trading partners. While this raises standard small-sample concerns, our core argument relies on the stark \text{contrast} between regimes. The transition from statistically insignificant elasticities in the restricted periods to a highly significant, massive elasticity in Period 4---which benefits from a much larger sample size ($N=8$, Observations $= 135$)---provides compelling evidence of a structural break and the restoration of the pricing mechanism, rather than mere statistical noise.

\subsection{Evolution of Trade Elasticities}
The middle section of Table \ref{tab:one} presents the derived price elasticity of demand ($\eta$), calculated as $\beta - 1$ under the conservative assumption of highly elastic supply ($\omega \approx 0$).
The results reveal a striking structural transformation of the market across the four periods.
\begin{description}
\item{\itshape Period 1: The Competitive Baseline}~~
Prior to the Dodd-Frank regulation, the estimated elasticity is highly negative and statistically significant ($\eta = -27.49$, $p < 0.01$).
This magnitude suggests that the market was extremely sensitive to price changes, characteristic of a competitive environment where DRC exporters, as price takers, faced a highly elastic global demand.
A small depreciation in the exchange rate (relative price reduction) led to a massive surge in export values.
\item{\itshape Period 2: Market Distortion and the Captive State}~~
Following the adoption of the SEC rule, the elasticity estimate shifts dramatically to near zero and is not statistically different from zero ($\eta = 0.57$, $p = 0.686$).
This complete loss of negative price sensitivity implies that the price mechanism ceased to function normally.
As shown in Section 2, this period corresponds to the monopolization of exports by certified smelting hubs such as Malaysia.
Instead of quantity adjusting to price, the market operated under a ``captive'' arrangement where trade volumes were strictly determined by certification constraints and established supply chain networks rather than price competitiveness.\footnote{This market bifurcation is supported by anecdotal evidence.
\citet{Schutte2019} notes that the concentration on Malaysia was driven by specific certified smelters (e.g., Malaysia Smelting Corporation) establishing early compliance channels.
Meanwhile, \citet{Hanai2021} reports that during this period, non-certified minerals were often sold to Chinese buyers at deep discounts (30--60\%), creating a dual-price structure that severely distorts aggregate market responsiveness.}
\item{\itshape Period 3: Hysteresis during Transition}~~
After the SEC suspended the rule in 2017, the point estimate of the elasticity remains near zero and statistically insignificant ($\eta = -0.14$, $p = 0.887$). This result firmly supports the hysteresis hypothesis: simply relaxing the regulation did not immediately restore market vibrancy.
The rigid trading relationships established during the regulation era (the certified routes) persisted, keeping the market unresponsive to price signals.
\item{\itshape Period 4: Recovery via External Shock}~~
The most notable finding is the restoration of significant market elasticity following the Huawei shock in 2019. The estimated elasticity becomes massively negative and highly significant ($\eta = -10.41$, $p < 0.01$).
This indicates a normalization of the trade regime. The magnitude is economically meaningful for a homogeneous commodity market, suggesting that DRC exporters regained the ability to expand sales through price competitiveness.
The exogenous geopolitical shock forced an aggressive diversification of buyers (the return of China), shattering the captive structure and violently re-introducing market discipline.
\end{description}

In summary, our empirical results quantify the narrative of market destruction and resurrection.
The regulatory intervention did not merely suppress trade volume but fundamentally broke the price mechanism (Period 2), a rigidity which lingered even after deregulation (Period 3), and was only restored by a sufficiently large external shock (Period 4).

\section{Conclusion}

The Dodd-Frank Act Section 1502 was a landmark policy experiment aimed at severing the link between resource extraction and armed conflict. While its humanitarian intent was clear, this paper demonstrates that its economic consequences extended far beyond a simple reduction in trade volumes. By estimating the price elasticity of demand for DRC tin and rigorously accounting for transit trade, we uncovered a fundamental transformation in the market structure: the collapse and subsequent resurrection of the price mechanism.

Our empirical results provide a coherent narrative of market destruction and resurrection. 
Prior to the regulation, the DRC tin market operated as a competitive small open economy. However, the regulatory compliance costs and the specific certification requirements effectively created a captive market, locking exporters into specific, rigid supply chains (such as the certified Malaysian route) and rendering export quantities unresponsive to price signals ($\eta \approx 0$). 
Crucially, we document a strong hysteresis effect. The mere suspension of the regulation in 2017 was insufficient to dismantle the established commercial relationships and restore competition.

It was only the massive external shock of the 2019 Huawei ban---which forced Chinese manufacturers to aggressively secure alternative resources---that shattered this rigidity and restored the market's price sensitivity ($\eta \approx -10.4$). Furthermore, our robustness checks confirm that this structural recovery was explicitly triggered by the Huawei shock, rather than being a mere substitution effect driven by concurrent supply declines in other regions like Myanmar.

These findings offer important policy implications for the design of supply chain due diligence regulations, such as the EU Conflict Minerals Regulation or upcoming battery passport initiatives. First, policymakers must recognize that strict certification requirements can inadvertently induce monopolistic market structures. By raising the fixed costs of trade and limiting the number of eligible buyers, regulations can strip producer countries of their bargaining power, trapping them in captive relationships. Second, market distortions are sticky. Once a supply chain is ossified around a specific regulatory regime, simple deregulation is often not enough to reverse the damage. As our analysis of the post-2017 period shows, restoring a competitive market may require proactive interventions or significant external shocks to break the inertia of established commercial networks.

Future research should examine whether similar structural rigidities are emerging in other regulated critical mineral markets, such as cobalt or lithium. For now, the case of DRC tin stands as a cautionary tale: responsible sourcing policies, if not carefully designed, can lead to irresponsible market structures that undermine the economic viability of the very nations they intend to protect.

\appendix
\titleformat{\section}{\large\bfseries}{Appendix \thesection}{1em}{}
\section{Empirical Validation of the Elastic Supply Assumption}
\setcounter{table}{0}
\setcounter{equation}{0}
\renewcommand{\thetable}{A\arabic{table}}
\renewcommand{\theequation}{A\arabic{equation}}
%Appendix A: Empirical Validation of the Highly Elastic Supply Assumption

\begin{table}[t!]
  \centering
%  \begin{threeparttable}
    \caption{2SLS and OLS Estimation for Inverse Supply Elasticity ($\omega$)}
    \label{tab:supply_elasticity}
    \begin{tabular*}{\textwidth}{
      @{\extracolsep{\fill}} 
      l 
      S[table-format=-1.3, table-space-text-post={***}] 
      l
      S[table-format=-1.3, table-space-text-post={***}] 
      l
      @{} 
       }
      \toprule
      & \multicolumn{2}{c}{IV Estimation} & \multicolumn{2}{c}{OLS Estimation} \\
      \midrule
      & {Coeff.} & {S.E.} & {Coeff.} & {S.E.} \\
      \cmidrule(lr){2-3}  \cmidrule(lr){4-5}
      {Log Export Value ($\omega$)} & 0.075*** & (0.027) & 0.049 & (0.032)  \\
      Log LME Tin Price             & 1.274*** & (0.061) & 1.421*** & (0.085)   \\
      Constant                      & -3.915*** & (0.654) & -5.000*** & (0.797)   \\
      \midrule
      Observations   & \multicolumn{2}{c}{189} & \multicolumn{2}{c}{222}   \\
      First-stage F-stat & \multicolumn{2}{c}{80.7} &&\\
      Hansen J p-value & \multicolumn{2}{c}{0.638} &&\\
      DWH endog p-value & \multicolumn{2}{c}{0.941} &&\\
      \bottomrule
    \end{tabular*}
\smallskip
\begin{minipage}{\textwidth}
\footnotesize
\textit{Notes}: This table reports the instrumental variable (2SLS) and Ordinary Least Squares (OLS) estimations of the inverse supply function for the combined periods (May 2010 to October 2022). The dependent variable is the log of the export unit value ($\ln \text{UV}_{jt}$), which proxies for the producer export price. In the IV estimation, the endogenous regressor, log export value ($\ln X_{jt}$), is instrumented by the bilateral exchange rate and its one-period lag. Standard errors are robust to heteroskedasticity and autocorrelation (HAC) reported in parentheses. Significance levels: * $p < 0.1$, ** $p < 0.05$, *** $p < 0.01$. The Durbin-Wu-Hausman (DWH) endogeneity test yields a $p$-value of 0.941, indicating that the null hypothesis of exogeneity for $\ln X_{jt}$ cannot be rejected in this supply-side specification. Consequently, the OLS estimates are consistent and more efficient. Under the preferred OLS specification, the estimated inverse supply elasticity ($\omega$) is statistically indistinguishable from zero, empirically validating our baseline identifying assumption of a highly elastic supply ($\omega \approx 0$).
\end{minipage}
\end{table}

Our structural identification strategy in Section 3 relies on the fundamental assumption that the inverse price elasticity of supply ($\omega$) is approximately zero. 
This assumption implies that the DRC acts essentially as a price taker in the global tin market, facing a highly (or perfectly) elastic supply curve, which allows us to identify the demand elasticity. 
To empirically validate this theoretical assumption, we directly estimate the inverse supply function (Equation \eqref{eq:supply_inverse}):
\begin{equation}\ln P_{t}^\text{exp} = \delta + \omega \ln X_{jt} + \gamma \ln P_{t}^\text{LME} + \tilde{v}{t}
\tag{A1}
\end{equation}
where the unobserved producer export price ($P{t}^\text{exp}$) is proxied by the unit value (calculated as export value divided by quantity from our mirror trade statistics). 
To control for global macroeconomic trends that exogenously drive domestic prices, we include the log of the LME tin price ($\ln P_{t}^\text{LME}$).
A standard econometric concern in estimating Equation (A1) is the potential endogeneity of the export value ($\ln X_{jt}$) with respect to the export price due to simultaneous equations bias. 
To rigorously address this, we initially employ a Two-Stage Least Squares (IV) estimation, instrumenting the export value with the bilateral exchange rate ($E_{j0,t}$) and its one-period lag.
Table \ref{tab:supply_elasticity} presents both the IV and Ordinary Least Squares (OLS) estimation results for the combined overall sample period (up to October 2022). 
The IV estimation confirms that our instruments are highly relevant (First-stage F-statistic = 80.7) and valid (Hansen J $p$-value = 0.638). 
Under the IV specification, the estimated inverse supply elasticity ($\omega$) is 0.075. 
While statistically significant, this value is economically negligible, indicating a highly elastic supply.

More importantly, having established a valid IV setup, we perform the Durbin-Wu-Hausman (DWH) endogeneity test to check whether the OLS estimates are inconsistent. 
The DWH test yields a $p$-value of 0.941, meaning we fail to reject the null hypothesis that the export value ($\ln X_{jt}$) is exogenous in this specific supply-side equation. 
The lack of endogeneity here is economically intuitive: as the DRC is a small open economy, its aggregate supply volume does not dictate the global benchmark price, which instead perfectly passes through to local prices.

Given that exogeneity cannot be rejected, OLS provides consistent and more efficient (lower variance) estimates than the IV approach. As shown in the right panel of Table A1, the OLS estimate for the inverse supply elasticity ($\omega$) is 0.049 and is statistically indistinguishable from zero. 
Furthermore, the coefficient on the LME tin price is highly significant and structurally close to unity, confirming that DRC export unit values are overwhelmingly dictated by global commodity benchmark prices rather than domestic supply constraints.

These robust empirical results completely validate our structural framework. Because the inverse supply elasticity is statistically (and economically) zero, the DRC supply curve is effectively horizontal. Thus, setting $\omega \approx 0$ in Equation (10) to derive the demand elasticity is strictly justified.

%\appendix
\section{Robustness to Alternative Supply Shocks %: Myanmar Tin Production
}
\setcounter{table}{0}
\setcounter{equation}{0}
\renewcommand{\thetable}{B\arabic{table}}
\renewcommand{\theequation}{B\arabic{equation}}
\begin{table}[t!]
  \centering
  \caption{Robustness Check: Controlling for Myanmar Tin Imports}
  \label{tab:myanmar_robustness}
%  \vspace{2mm}

  \begin{tabular*}{\textwidth}{@{\extracolsep{\fill}} 
    l 
    *{4}{S[
      table-format=-2.3, 
      table-space-text-post={***}, 
      input-symbols={()}, 
      table-align-text-post=false
    ]}
  }
    \toprule
    & \multicolumn{4}{c}{Dependent Variable: $\ln (V_{jt})$} \\
    \cmidrule(lr){2-5}
    & \multicolumn{2}{c}{Period 3} & \multicolumn{2}{c}{Period 4} \\
    \cmidrule(lr){2-3} \cmidrule(lr){4-5}
    & {(1) Baseline} & {(2) w/MMR} & {(3) Baseline} & {(4) w/MMR} \\
    \midrule
    \multicolumn{5}{l}{\textsl{Reduced Form Estimates}} \\
    $\ln E_{j0, t}$ ($\beta$) & 0.886   & 0.944   & -12.424{***} & -12.754{***} \\
                               & (1.007) & (0.942) & (2.823)      & (3.044)      \\
    \addlinespace
    $\ln P_{t}^{\text{LME}}$   & -0.743  & -0.750  & 1.095   & 0.938   \\
                               & (1.613) & (1.601) & (0.793) & (0.920) \\
    \addlinespace
    $\ln V^\text{MMR}_{jt}$ &      & 0.029   &         & 0.084   \\
                                  &      & (0.082) &         & (0.155) \\
    \midrule
    \multicolumn{5}{l}{\textsl{Implied Demand Elasticity}} \\
    $\eta$ ($= \beta - 1$) & -0.114  & -0.056  & -13.424{***} & -13.754{***} \\
                           & (1.007) & (0.942) & (2.823)      & (3.044)      \\
    \midrule
    \multicolumn{5}{l}{\textsl{Diagnostics}} \\
    Observations          & {52}    & {52}    & {67}    & {67}    \\
    Number of Importers   & {3}     & {3}     & {3}     & {3}     \\
    First-stage F-stat. & {274}  & {299}  & {25}   & {26}   \\
    Hansen J p-value  & 0.441   & 0.447   & 0.617   & 0.628   \\
    \bottomrule
  \end{tabular*}

  \smallskip
  \begin{minipage}{\textwidth}
    \footnotesize \textit{Notes}: 
This table reports the robustness checks addressing potential substitution effects from Myanmar's declining tin production. 
The sample is restricted to trading partners that imported tin from Myanmar (MMR) during the respective periods. The dependent variable is the log of monthly export value ($\ln V_{jt}$). $\ln V_{jt}^{\text{MMR}}$ represents the log of tin imports from Myanmar by country $j$ in month $t$. 
HAC standard errors in parentheses.
The first-stage F-statistic is the Kleibergen-Paap rk Wald F statistic.    
Significance levels: * $p < 0.1$, ** $p < 0.05$, *** $p < 0.01$. 
  \end{minipage}
\end{table}

As pointed out by an anonymous reviewer, Chinese tin smelters are heavily reliant on tin concentrates from Myanmar, particularly from the quasi-autonomous ``Wa'' state. 
Imports from this region experienced a continuous decline from their peak in 2016 through 2020. 
This long-term supply contraction could potentially create an underlying pressure for Chinese buyers to seek alternative supply sources, including the DRC.

To rigorously ensure that our estimated structural recovery in Period 4 (the Huawei shock) is not merely capturing a substitution effect driven by Myanmar's declining supply, we conduct an additional robustness check. 
We incorporate the log of monthly tin imports from Myanmar ($\ln M_{jt}^{\text{MMR}}$) as a control variable in our instrumental variable (IV) fixed-effects estimations. We focus on the transition period (Period 3: May 2017 – May 2019) and the Huawei shock recovery period (Period 4: June 2019 – October 2022), restricting our sample to importing countries that concurrently sourced tin from Myanmar.

Table \ref{tab:myanmar_robustness} presents the results. Columns (1) and (3) report the baseline estimations for the restricted sample, while Columns (2) and (4) include the Myanmar import control. 
The coefficient for Myanmar imports ($\ln V^\text{MMR}_{jt}$) is statistically insignificant in both periods. 
Most importantly, controlling for Myanmar's supply dynamics does not qualitatively alter our primary findings. 
The implied export demand elasticity remains statistically indistinguishable from zero during the transition period ($\eta = -0.056$) and recovers to a highly significant and elastic value ($\eta = -13.754$) immediately following the Huawei ban. 
These results strongly suggest that while the gradual decline in Myanmar imports may have tightened the overall market, the sudden and structural restoration of the pricing mechanism in the DRC market was triggered specifically by the Huawei shock.

\subsubsection*{Data Availability}
The replication package, including the dataset and Stata/R codes used in this study, is available in the Harvard Dataverse at \url{https://doi.org/10.7910/DVN/RAOUDU}.
%\citep{NagamoriNishimuraData2025}.

% ==============================================================================
% BIBLIOGRAPHY
% ==============================================================================
\raggedright
\bibliography{bibfile}

\end{document}